\documentstyle[multicol,aps,psfig]{revtex}
\renewcommand{\narrowtext}{\begin{multicols}{2}
\global\columnwidth20.5pc\noindent}
\renewcommand{\widetext}{\end{multicols}
\global\columnwidth42.5pc}
\begin{document}
\draft
\preprint{Ag2060, PLA9533}
\title{Nuclear spin relaxation in ordered bimetallic chain compounds}
\author{Shoji Yamamoto}
\address{Department of Physics, Okayama University,
         Tsushima, Okayama 700-8530, Japan}
\date{Received 18 November 1999}
\maketitle
\begin{abstract}
A theoretical interpretation is given to recent proton spin
relaxation-time ($T_1$) measurements on
NiCu(C$_7$H$_6$N$_2$O$_6$)(H$_2$O)$_3$$\cdot$2H$_2$O,
which is an ideal one-dimensional ferrimagnetic Heisenberg model
system of alternating spins $1$ and $\frac{1}{2}$.
The relaxation rate $T_1^{-1}$ is formulated in temrs of the
spin-wave theory and is evaluated by the use of a quantum Monte Carlo
method.
Calculations of the temperature and applied-field ($H$) dependences
of $T_1^{-1}$ are in total agreement with the experimental findings.
$T_1$ behaves as $T_1^{-1}\propto H^{-1/2}$, which turns out an
indirect observation of the quadratic dispersion relations dominating
the low-energy physics of quantum ferrimagnets.
\end{abstract}
\narrowtext

   Recently considerable attention has been directed to
ferrimagnetic mixed-spin chains.
The simplest but practical example of a quantum ferrimagnet is two
kinds of spins $S$ and $s$ alternating on a ring with
antiferromagnetic exchange coupling between nearest neighbors.
Because of the noncompensating sublattice magnetizations, this system
exhibits ferrimagnetically degenerate ground states \cite{Lieb49}.
Thus, in contrast with ferromagnets and antiferromagnets,
ferrimagnets show the ground-state excitations of dual aspect
\cite{Yama10}.
The elementary excitations of ferromagnetic features, reducing the
ground-state magnetization, form a gapless dispersion relation,
whereas those of antiferromagnetic features, enhancing the
ground-state magnetization, are gapped from the ground state.
The two distinct low-lying excitations result in the unique thermal
behavior which contains both ferromagnetic and antiferromagnetic
aspects \cite{Dril13,Pati94,Yama08,Wu57}.
Quantum ferrimagnets in a magnetic field provide further interesting
topics such as the double-peak structure of the specific heat
\cite{Mais08} and quantized plateaux in the ground-state
magnetization curves \cite{Kura62,Saka,Hagi}.

   The first ferrimagnetic one-dimensional compound,
MnCu(dto)$_2$(H$_2$O)$_3$$\cdot$$4.5$H$_2$O
($\mbox{dto}=\mbox{dithiooxalato}=\mbox{S}_2\mbox{C}_2\mbox{O}_2$),
was synthesized by Gleizes and Verdaguer \cite{Glei27}.
Their pioneering efforts, combined with semiclassical but
enlightening calculations, successfully characterized the
one-dimensional ferrimagnetic behavior \cite{Verd44}.
Another example of an ordered bimetallic chain,
MnCu(pba)(H$_2$O)$_3$$\cdot$$2$H$_2$O
($\mbox{pba}=1,3\mbox{-propylenebis(oxamato)}
 =\mbox{C}_7\mbox{H}_6\mbox{N}_2\mbox{O}_6$),
was provided by Pei {\it et al.} \cite{Pei38} and turned out to
exhibit more pronounced one dimensionality.
Kahn {\it et al.} \cite{Kahn82} further synthesized the family of
related compounds systematically, focusing on the problem of the
crystal engineering of a molecule-based ferromagnet$-$the assembly of
the highly magnetic molecular entities within the crystal lattice in
a ferromagnetic fashion.
There also appeared an idea \cite{Cane56} that the alternating
magnetic centers do not need to be metal ions but may be organic
radicals.

   In comparison with the accumulated chemical knowledge and
well-revealed static properties introduced above, the dynamic
properties of quantum ferrimagnets have much less been studied so
far.
To the best of our knowledge, it was not until quite recently that
the dynamic structure factors were calculated \cite{Yama11}, whereas
any direct observation of the dispersion curves is not yet so
successful, for instance, as that \cite{Ma71} for the Haldane
antiferromagnets \cite{Hald64}.
While Caneschi {\it et al.} \cite{Cane56} performed
electron-paramagnetic-resonance measurements on novel ferrimagnetic
compounds, which consist of alternating metal ions and stable organic
radicals, they focused their analyses on the relation between the
spectra and short-range correlation effects.
In such circumstances, the proton spin relaxation time $T_1$ has
quite recently been measured \cite{Fuji} for the
spin-$(1,\frac{1}{2})$ ferrimagnetic Heisenberg chain compound,
NiCu(pba)(H$_2$O)$_3$$\cdot$2H$_2$O, where the dependence of the
relaxation rate $T_1^{-1}$ on the applied magnetic field $H$ looks
like $T_1^{-1}\propto H^{-1/2}$, though the authors' semiquantitative
approximate argument has reached a logarithmic behavior,
$T_1^{-1}\sim{\rm ln}H$, rather than the observations.
Motivated by this stimulative experiment, we here discuss the
nuclear spin relaxation peculiar to ordered bimetallic chain
compounds in connection with their energy structure.
Formulating the relaxation process based on the spin-wave theory, we
perform numerical evaluation of $T_1^{-1}$.
Investigating the dependences of $T_1^{-1}$ on temperature and the
applied field, we demonstrate that the $T_1$ measurement is nothing
but an indirect observation of the characteristic elementary
excitations of quantum ferrimagnets.

   The ferrimagnetic material  NiCu(pba)(H$_2$O)$_3$$\cdot$2H$_2$O
\cite{Pei38} consists of ordered bimetallic chains with alternating
octahedral Ni$^{2+}$ and square-pyramidal Cu$^{2+}$ ions bridged by
oxamato groups.
The chain runs along the $b$-axis of the orthorhombic lattice whose
space group is $Pnma$ and is described by the Hamiltonian
\begin{equation}
   {\cal H}
      =J\sum_{j=1}^N
        \left(
         \mbox{\boldmath$S$}_{j} \cdot \mbox{\boldmath$s$}_{j}
        +\mbox{\boldmath$s$}_{j} \cdot \mbox{\boldmath$S$}_{j+1}
        \right)
      -g\mu_{\rm B} H\sum_{j=1}^N(S_j^z+s_j^z)\,,
   \label{E:H}
\end{equation}
where we have set the $g$ factors of spins $S=1$ and $s=\frac{1}{2}$
both equal to $g$ because the difference between them amounts to at
most several per cent of themselves in practice \cite{Hagi09}.
The high-temperature susceptibility measurements suggest that
$J\simeq 121[\mbox{K}]$.
The one dimensionality is well exhibited above about $7[\mbox{K}]$.
Due to the energy-conservation requirement for the
electronic-nuclear spin system, the direct (single-magnon) process
is of little significance but the Raman (two-magnon) process plays a
leading role in the nuclear spin-lattice relaxation \cite{Beem59}.
Neglecting the higher-order relaxation process, the relaxation rate
is generally represented as
\widetext
\begin{equation}
   \frac{1}{T_1}
    =\frac{4\pi(g\mu_{\rm B}\hbar\gamma_{\rm N})^2}
          {\hbar\sum_n{\rm e}^{-E_n/k_{\rm B}T}}
     \sum_{n,m}{\rm e}^{-E_n/k_{\rm B}T}
     \big|
      \langle m|{\scriptstyle\sum_j}(A_j^zS_j^z+a_j^zs_j^z)|n\rangle
     \big|^2
     \,\delta(E_m-E_n-\hbar\omega_{\rm N})\,,
\label{E:T1def}
\end{equation}
\narrowtext
where
$A_j^z$ and $a_j^z$ are the dipolar coupling constants between
the proton and electron spins in the $j$th unit cell,
$\omega_{\rm N}\equiv\gamma_{\rm N}H$ is the Larmor
frequency of $^1$H with $\gamma_{\rm N}$ being the gyromagnetic
ratio, and the summation $\sum_n$ is taken over all the eigenstates
of $|n\rangle$ with energy $E_n$.
Let us introduce the bosonic spin-deviation operators via
$S_j^+=\sqrt{2S}\,a_j$,
$S_j^z=S-a_j^\dagger a_j$,
$s_j^+=\sqrt{2s}\,b_j^\dagger$, and
$s_j^z=-s+b_j^\dagger b_j$.
The Bogoliubov transformation
\begin{equation}
   \left.
   \begin{array}{c}
      \alpha_k=\mbox{cosh}\theta_k\ a_k
              +\mbox{sinh}\theta_k\ b_k^\dagger\,,\\
      \beta_k =\mbox{sinh}\theta_k\ a_k^\dagger
              +\mbox{cosh}\theta_k\ b_k\,,
   \end{array}
   \right.
   \label{E:Bogoliubov}
\end{equation}
with
\begin{equation}
   \left.
   \begin{array}{c}
      a_k=\displaystyle{\frac{1}{\sqrt{N}}}
          \sum_j{\rm e}^{ {\rm i}k(j-\frac{1}{4})}a_j\,,\\
      b_k=\displaystyle{\frac{1}{\sqrt{N}}}
          \sum_j{\rm e}^{-{\rm i}k(j+\frac{1}{4})}b_j\,,
   \end{array}
   \right.
\end{equation}
and
\begin{equation}
   {\rm tanh}(2\theta_k)
    =\frac{2\sqrt{Ss}}{S+s}
     {\rm cos}\Bigl(\frac{k}{2}\Bigr)\,,
\end{equation}
diagonalizes the Hamiltonian (\ref{E:H}), discarding the constant
terms, as \cite{Pati94,Breh21}
\begin{equation}
   {\cal H}=
      \sum_k
      \left(
       \omega_{k}^-\alpha_k^\dagger\alpha_k
      +\omega_{k}^+\beta_k^\dagger \beta_k
      \right)\,,
   \label{E:SWH}
\end{equation}
where
$\alpha_k^\dagger$ and $\beta_k^\dagger$ are the creation operators
of the ferromagnetic and antiferromagnetic spin waves of momentum
$k$ whose dispersion relations are given by
\begin{eqnarray}
   \omega_{k}^\pm
   &=&J\sqrt{(S-s)^2+4Ss\sin^2(k/2)}
   \nonumber \\
   &\pm&(S-s)J\mp g\mu_{\rm B}H\,,
   \label{E:SWDSP}
\end{eqnarray}
with twice the lattice constant being taken as unity.
In terms of the spin waves, the relaxation rate (\ref{E:T1def}) is
expressed as
\begin{eqnarray}
   \frac{1}{T_1}
   &=&\frac{4\pi\hbar}{N^2}(g\mu_{\rm B}\gamma_{\rm N})^2
      \sum_{k,q}\sum_{\sigma=\pm}
      \delta(\omega_{k+q}^\sigma-\omega_k^\sigma-\hbar\omega_{\rm N})
   \nonumber \\
   &\times&
   \bigl[
    (A_q^z{\rm cosh}\theta_{k+q}{\rm cosh}\theta_k)^2
    n_k^\sigma(n_{k+q}^\sigma+1)
   \nonumber \\
   &&\!\! +
    (a_q^z{\rm sinh}\theta_{k+q}{\rm sinh}\theta_k)^2
    n_{k+q}^\sigma(n_k^\sigma+1)
   \nonumber \\
   &&\!\! -
    2A_q^z a_q^z({\rm cosh}\theta_k{\rm sinh}\theta_k)^2
    n_k^\sigma(n_k^\sigma+1)
   \bigr]\,,
   \label{E:T1SW}
\end{eqnarray}
where
$n_k^-\equiv\langle\alpha_k^\dagger\alpha_k\rangle$ and
$n_k^+\equiv\langle \beta_k^\dagger\beta_k \rangle$ are the thermal
averages of the numbers of the spin waves at a given temperature,
and
$A_q^z=\sum_j{\rm e}^{{\rm i}q(j-1/4)}A_j^z$ and
$a_q^z=\sum_j{\rm e}^{{\rm i}q(j+1/4)}a_j^z$
are the Fourier components of the hyperfine coupling constants.
Taking into account the significant difference between the electronic
and nuclear energy scales ($\hbar\omega_{\rm N}\alt 10^{-5}J$), Eq.
(\ref{E:T1SW}) ends in
\begin{eqnarray}
   \frac{1}{T_1}
   &=&\frac{4\hbar}{NJ}(g\mu_{\rm B}\gamma_{\rm N})^2
      \sum_k\frac{S-s}{\sqrt{(Ssk)^2+2(S-s)Ss\hbar\omega_{\rm N}/J}}
   \nonumber \\
   &\times&
   \bigl[
    (A^z{\rm cosh}^2\theta_k-a^z{\rm sinh}^2\theta_k)^2
    n_k^-(n_k^- +1)
   \nonumber \\
   &&\!\! +
    (A^z{\rm sinh}^2\theta_k-a^z{\rm cosh}^2\theta_k)^2
    n_k^+(n_k^+ +1)
   \bigr]\,,
   \label{E:T1final}
\end{eqnarray}
where
we have assumed little $k$-dependence of $A_q^z$ and
$a_q^z$, and thus replaced $A_{q=-2k}^z$ and $a_{q=-2k}^z$ by
$A^z\equiv A_{q=0}^z$ and $a^z\equiv a_{q=0}^z$, respectively.

\begin{figure}
\begin{flushleft}
\ \ \mbox{\psfig{figure=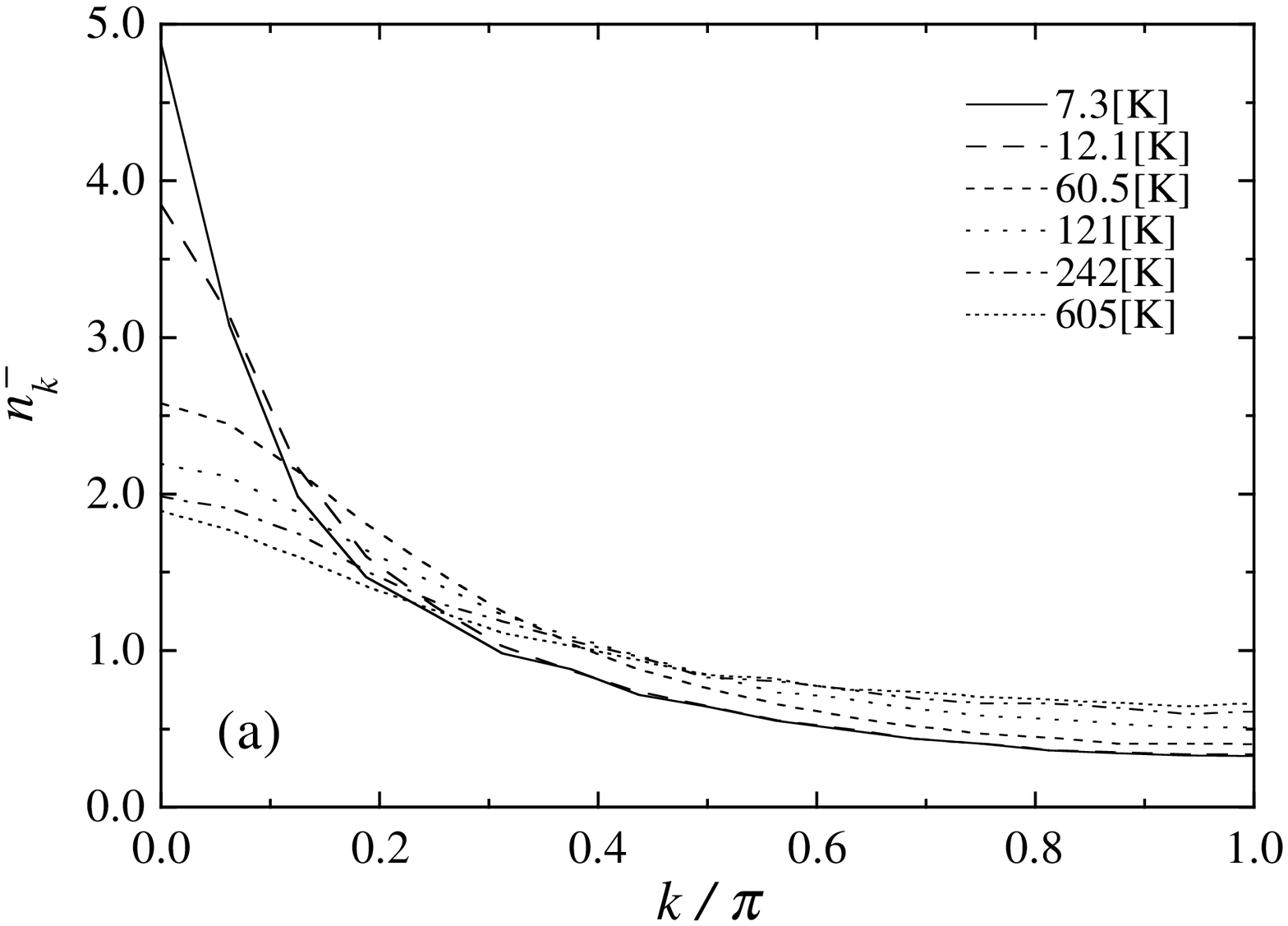,width=80mm,angle=0}}
\vskip 0mm
\ \ \mbox{\psfig{figure=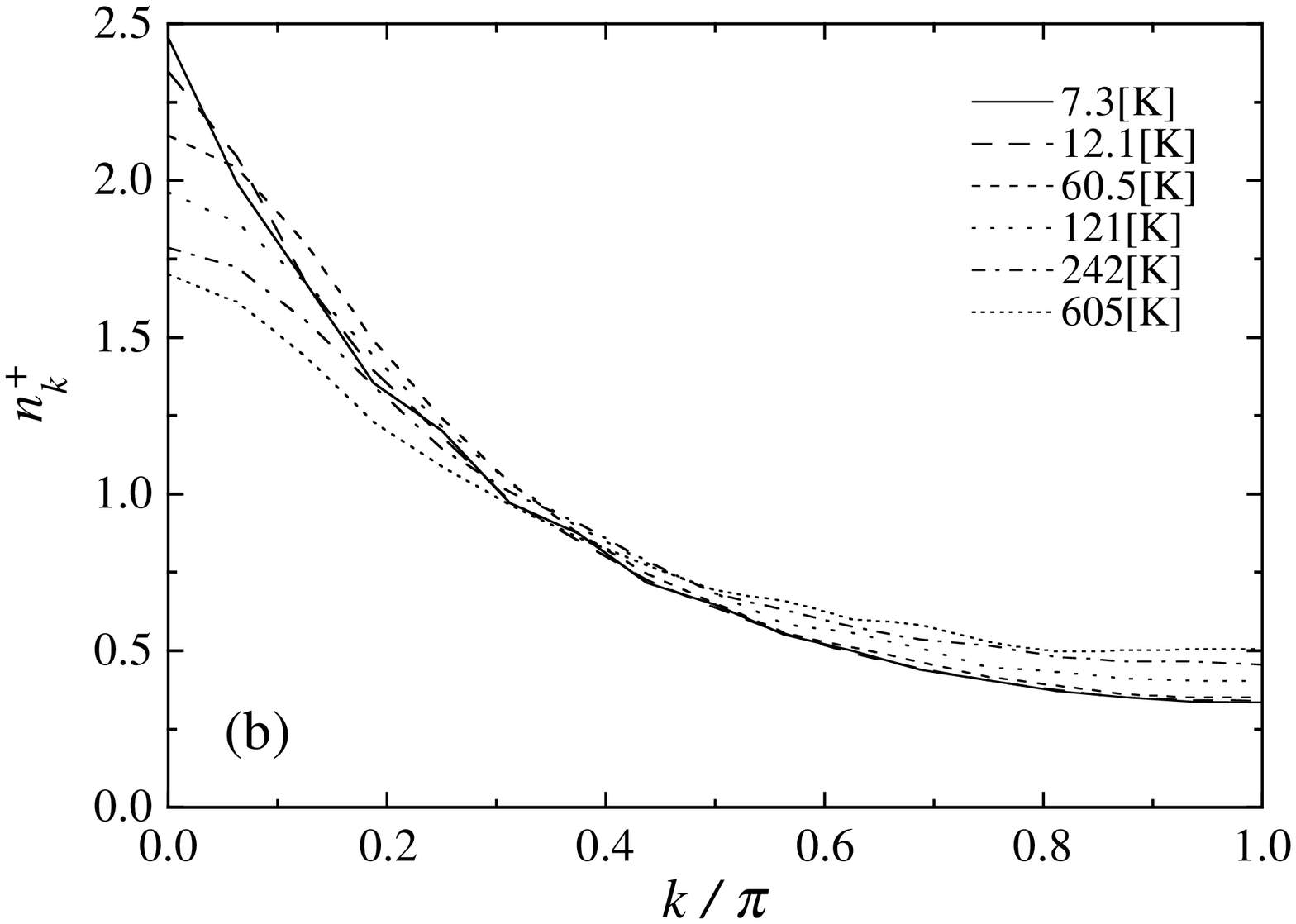,width=80mm,angle=0}}
\end{flushleft}
\vskip 0mm
\caption{Quantum Monte Carlo estimates of the ferromagnetic (a) and
         antiferromagnetic (b) static structure factors at various
         values of temperature.}
\label{F:nk}
\end{figure}

   Now we have an idea of evaluating the thermal averages $n_k^\pm$
by a quantum Monte Carlo method.
Though the divergence of the ground-state sublattice magnetization,
which plagues the spin-wave treatment of low-dimensional
antiferromagnets, does not persist in our model, $n_k^\pm$ still
diverge as temperature increases as far as we naively estimate them
with the noncompact bosonic Hamiltonian (\ref{E:SWH}).
On the other hand, we are fully allowed to rely upon the dispersion
relations (\ref{E:SWDSP}) themselves even at finite temperatures,
because we can efficiently modify the spin-wave thermodynamics
\cite{Yama08} so as to describe the higher-temperature behavior
introducing an additional constraint on the magnetization but
preserving the linearized spin-wave dispersions $\omega_k^\pm$.
Thus, depending on the Bogoliubov transformation (\ref{E:Bogoliubov})
but avoiding the direct estimation of the boson numbers, we perform
the Monte Carlo sampling for the relevant spin operators
\begin{equation}
   \left.
   \begin{array}{l}
     \alpha_k
     ={\displaystyle\frac{{\rm cosh}\theta_k}{\sqrt{2N}}}
      {\displaystyle\sum_j}{\rm e}^{ {\rm i}k(j-\frac{1}{4})}S_j^+
     +{\displaystyle\frac{{\rm sinh}\theta_k}{\sqrt{ N}}}
      {\displaystyle\sum_j}{\rm e}^{ {\rm i}k(j+\frac{1}{4})}s_j^+
     \,,\\
     \beta_k
     ={\displaystyle\frac{{\rm sinh}\theta_k}{\sqrt{2N}}}
      {\displaystyle\sum_j}{\rm e}^{-{\rm i}k(j-\frac{1}{4})}S_j^-
     +{\displaystyle\frac{{\rm cosh}\theta_k}{\sqrt{ N}}}
      {\displaystyle\sum_j}{\rm e}^{-{\rm i}k(j+\frac{1}{4})}s_j^-
     \,,\\
   \end{array}
   \right.
\end{equation}
with the original compact Hamiltonian (\ref{E:H}).
While we base the relaxation process on the spin-wave excitations, we
take grand-canonical averages within the original system.
Since the applied field $H$ is so small as to satisfy
$g\mu_{\rm B}H\alt 10^{-2}J$, we neglect the Zeeman term of the
Hamiltonian (\ref{E:H}) in the numerical treatment.
The thus-calculated structure factors $n_k^\pm$ are shown in Fig.
\ref{F:nk}.
$n_k^\pm$ both have their peaks at $k=0$ in the reduced Brillouin
zone, which reflects the combination of the ferromagnetic and
antiferromagnetic features.
The momentum dependences of $n_k^\pm$ are indeed consistent with the
exact calculations \cite{Yama11} of the dynamic structure factors.
The temperature dependences of $n_k^\pm$ well support the
characteristic Schottky-like peak of the specific heat \cite{Yama08},
where the ferromagnetic structure is rapidly smeared out as
temperature increases, while the antiferromagnetic one still persists
at rather high temperatures.
Therefore the present analysis is promising.
The dipolar coupling is quite sensitive to the location of the proton
because the coupling strength is proportional to $r^{-3}$ with $r$
being the distance between the interacting proton and electron spins.
In the experiment on
NiCu(pba)(H$_2$O)$_3$$\cdot$2H$_2$O,
it has been demonstrated \cite{Fuji} that the protons mainly
contributing to $T_1^{-1}$ lie in the pba groups rather than the
H$_2$O molecules and are therefore all located beside Cu ions.
Hence we here reasonably neglect $A^z$ in comparison with $a^z$.
Setting $a^z$ equal to $6.12\times 10^{-3}[\mbox{\AA}^{-3}]$ with
$g=2$, the calculations successfully reproduce the observations.
\vskip 4mm
\vskip 0mm
\begin{figure}
\begin{flushleft}
\ \ \mbox{\psfig{figure=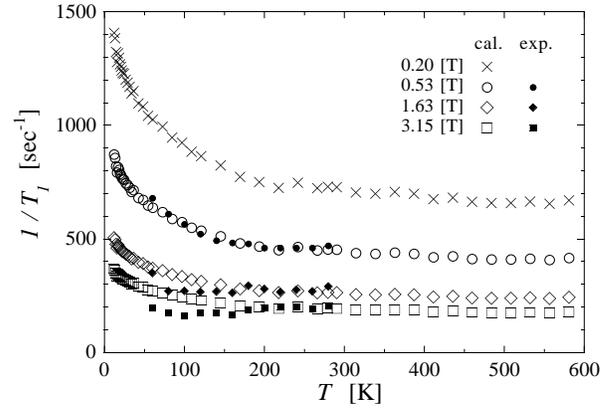,width=80mm,angle=0}}
\end{flushleft}
\vskip 0mm
\caption{Temperature dependence of the proton spin relaxation rate as
         a function of the applied magnetic field.
         The measurements (exp.) and calculations (cal.) are
         indicated by closed and open symbols, respectively.}
\label{F:T1T}
\end{figure}
\vskip 2mm
\begin{figure}
\begin{flushleft}
\ \ \mbox{\psfig{figure=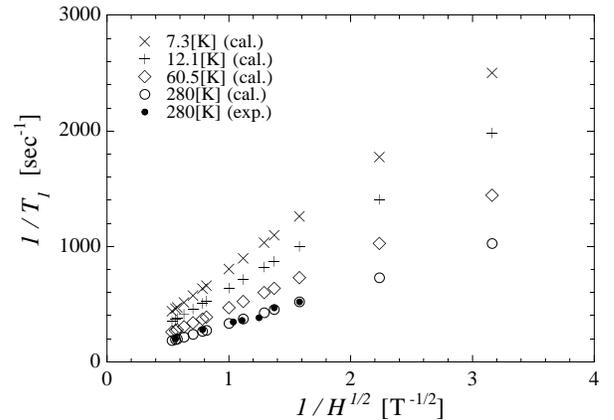,width=80mm,angle=0}}
\end{flushleft}
\vskip 0mm
\caption{Field dependence of the proton spin relaxation rate as a
         function of temperature.
         Closed circles and the others indicate the measurements
         (exp.) and calculations (cal.), respectively.}
\label{F:T1H}
\end{figure}
\vskip 2mm

   We show in Fig. \ref{F:T1T} the thus-obtained relaxation rate as
a function of temperature together with the observations.
The calculations are generally in good agreement with the
experimental findings.
The slight difference between them at low temperatures under
intermediate fields may be attributed to the uncertainty in the
experimental treatment, where the linewidth relatively broadens and
the recovery of the spin echo looks like a double-exponential curve.
Although the static susceptibility $\chi$ times temperature, which
is closely related with the relaxation rate \cite{Hone65}, shows a
round minimum at about $70[\mbox{K}]$ \cite{Pei38,Hagi09}, $T_1^{-1}$
monotonically decreases as $T$ increases.
This is, however, convincing if we bear the temperature dependences
of $n_k^\pm$ in mind.
Let us go back to Fig. \ref{F:nk}.
The peak of the antiferromagnetic static structure factor, which
usually appears at the zone boundary in mono-spin chains, now shifts
to the zone center due to the double periodicity in the present
system.
With the increase of temperature, the structure factor generally
decreases in the vicinity of its peaks, while it increases in its
lower slopes.
Thus, due to the predominance of the $k=0$ component in the summation
(\ref{E:T1final}), $T_1^{-1}$ ends up as a decreasing function of
temperature.
The increase of the applied field, having an effect of reducing the
predominance of the $k=0$ component, relaxes the pronounced
decreasing behavior of $T_1^{-1}$ as a function of temperature and
would possibly allow $T_1^{-1}$ to show a reversed but weak
temperature dependence at high temperatures.

   Another interesting observation is the field dependence of the
relaxation rate shown in Fig. \ref{F:T1H}.
$T_1^{-1}$ looks almost linear with respect to $H^{-1/2}$, which also
suggests that only the magnons of $k\simeq 0$ effectively contribute
to the relaxation process.
We stress that the present field dependence comes from the
energy-conservation requirement $\delta(E_m-E_n-\hbar\omega_{\rm N})$
in Eq. (\ref{E:T1def}), that is, from the quadratic dispersion
relations of the elementary excitations.
In this context, the spin-diffusion effect \cite{Hone65} may be
mentioned, which also results in a similar field dependence of
$T_1^{-1}$,
\begin{equation}
   \frac{1}{T_1}=P+\frac{Q}{\sqrt{H}}\,,
   \label{E:T1FD}
\end{equation}
where the constant $P$ arises from longitudinal spin fluctuations,
whereas the field-dependent term originates from transverse spin
fluctuations which are diffusive at high temperatures and long times.
For (CH$_3$)$_4$NMnCl$_3$ \cite{Bouc98} and
LiV$_2$O$_5$\cite{Fuji45}, which are both well known to be
spin-diffusive Heisenberg chain compounds, $P/Q$ was estimated to be
$0.57[\mbox{T}^{-1/2}]$ at $300[\mbox{K}]$ with $J= 13[\mbox{K}]$ and
$1.53[\mbox{T}^{-1/2}]$ at $200[\mbox{K}]$ with $J=308[\mbox{K}]$,
respectively.
Here we observe a much smaller value, $P/Q=0.05[\mbox{T}^{-1/2}]$ at
$280[\mbox{K}]$, which is much more consistent with the present
analysis.
Therefore, the present observations, which look like
$T_1^{-1}\propto H^{-1/2}$, should be attributed to the quadratic
dispersion relations peculiar to Heisenberg ferrimagnets.
We learn that the longitudinal spin fluctuations dominate the
nuclear spin relaxation in ordered bimetallic chain compounds.

   The present compound, NiCu(pba)(H$_2$O)$_3$$\cdot$2H$_2$O, is
reasonably described by the isotropic Heisenberg Hamiltonian.
There exist a series of family compounds
\cite{Glei27,Verd44,Pei38,Kahn82} and thus the prototypical
ferrimagnetic behavior \cite{Yama24} could potentially be elucidated
as a function of the constituent spins.
The consideration of anisotropy effects may also be interesting from
the practical point of view.
For the anisotropic Hamiltonian
\widetext
\begin{equation}
   {\cal H}
   =J\sum_j
    \bigl[
     (\mbox{\boldmath$S$}_{j}\cdot\mbox{\boldmath$s$}_{j})_\alpha
    +(\mbox{\boldmath$s$}_{j}\cdot\mbox{\boldmath$S$}_{j+1})_\alpha
    \bigr]
   +\sum_j
    \bigl[
     D_S(S_j^z)^2+D_s(s_j^z)^2
    \bigr]
   -g\mu_{\rm B}H\sum_j(S_j^z+s_j^z)\,,
   \label{E:Hani}
\end{equation}
with
$(\mbox{\boldmath$S$}\cdot\mbox{\boldmath$s$})_\alpha
 =S^xs^x+S^ys^y+\alpha S^zs^z$,
the relaxation rate (\ref{E:T1final}) is replaced by
\begin{eqnarray}
   \frac{1}{T_1}
   &=&\frac{4\hbar}{NJ}(g\mu_{\rm B}\gamma_{\rm N})^2
      \sum_k
      \frac{f_{S,s}(\alpha,D_S,D_s)}
           {\sqrt{(Ssk)^2
            +2Ssf_{S,s}(\alpha,D_S,D_s)\hbar\omega_{\rm N}/J}}
   \nonumber \\
   &\times&
   \bigl[
    (A^z{\rm cosh}^2\theta_k-a^z{\rm sinh}^2\theta_k)^2
    n_k^-(n_k^- +1)
   +(A^z{\rm sinh}^2\theta_k-a^z{\rm cosh}^2\theta_k)^2
    n_k^+(n_k^+ +1)
   \bigr]\,,
   \label{E:T1ani}
\end{eqnarray}
where
\begin{equation}
   f_{S,s}(\alpha,D_S,D_s)
   =\sqrt{[\alpha(S+s)-(SD_S+sD_s)/J]^2-4Ss}\,.
\end{equation}
\narrowtext
Even if any anisotropy of the easy-axis type is introduced, the
small-momentum quadratic dispersion remains qualitatively unchanged
and thus the characteristic field dependence of the relaxation rate,
$T_1^{-1}\propto H^{-1/2}$, should still be observed, where the slope
$\partial T_1^{-1}/\partial H^{-1/2}$ gives semiquantitative
information on the anisotropy.
However, the expression (\ref{E:T1ani}) is no more valid under the
easy-plane-type anisotropy.
The model indeed turns critical for $\alpha<1$, showing linear
dispersion relations \cite{Alca67}.
Therefore ferrimagnetic compounds with anisotropic interactions of
the easy-plane type should exhibit no field dependence of $T_1^{-1}$.
Thus the nuclear magnetic resonance is all the more efficient to
investigate the low-energy structure in the present system.
We really hope that the present successful interpretation of the
pioneering experiments will stimulate and accelerate further study
on this fascinating system.
The developed calculations under various constituent spins and
geometric parameters will be presented elsewhere \cite{Yama}.

   The author is grateful to Dr. N. Fujiwara for his communication on
the NMR measurements prior to publication.
This work was supported by the Japanese Ministry of Education,
Science, and Culture, and by the Sanyo-Broadcasting Foundation for
Science and Culture.
The numerical computation was done in part using the facility of the
Supercomputer Center, Institute for Solid State Physics, University of
Tokyo.

\widetext
\end{document}